%% file: main.tex
\documentclass[runningheads]{llncs}

\usepackage[T1]{fontenc}
\usepackage{xspace}
\usepackage{url}

\usepackage{hyperref}
\usepackage{makecell}
\usepackage{amsmath}
\usepackage{tikz}
\usepackage{enumitem}

\usepackage{caption}
\usepackage{subcaption}

\usepackage{listings}
\usepackage{mdframed}
\usepackage[capitalize]{cleveref}

\usepackage[section]{placeins}

\usepackage{graphicx}

\usepackage{color}

\newcommand{\mambov}{MAMBO{\bfseries--}\hspace{-.05em}V}
\newcommand{\mambo}{MAMBO}
\newcommand{\mw}{Microwalk}
\newcommand{\rv}{RISC{\bfseries--}\hspace{-.05em}V}
\newcommand{\arm}{AArch32}
\newcommand{\aarch}{AArch64}
\newcommand{\ct}{constant-time}
\newcommand{\nct}{non-constant-time}
\newcommand{\sic}{side-channel} %
\newcommand{\scs}{side-channels}
\newcommand{\clanguage}{C}
\newcommand{\cpplanguage}{{C\nolinebreak[4]\hspace{-.05em}\raisebox{.3ex}{\small ++}}}

\newcommand{\openssl}{OpenSSL}
\newcommand{\wolfssl}{WolfSSL}
\newcommand{\nettle}{GNU Nettle}
\newcommand{\scl}{SCL}
\newcommand{\mbedtls}{Mbed TLS}
\newcommand{\libsodium}{libsodium}

\newcommand{\gmp}{GMP}

\newcommand{\aesecb}{AES-ECB}

\newcommand{\aesgcm}{AES-GCM}
\newcommand{\chacha}{ChaCha20-Poly1305}
\newcommand{\edsig}{Ed25519}
\newcommand{\ecdsa}{ECDSA}

\usepackage{wasysym} %

\begin{document}

\title{\mambov{}: Dynamic Side-Channel Leakage\\Analysis on \rv{}}

\author{Jan Wichelmann \and Christopher Peredy \and Florian Sieck \and Anna Pätschke \and Thomas Eisenbarth}

\authorrunning{J. Wichelmann et al.}

\institute{University of Lübeck, Lübeck, Germany
\email{\{j.wichelmann,c.peredy,florian.sieck,a.paetschke,thomas.eisenbarth\}\\@uni-luebeck.de}}
\maketitle              %
\begin{abstract}

\rv{} is an emerging technology, with applications ranging from embedded devices to high-performance servers.
Therefore, more and more security-critical workloads will be conducted with code that is compiled for \rv{}.
Well-known microarchitectural \sic{} attacks against established platforms like x86 apply to \rv{} CPUs as well.
As \rv{} does not mandate any hardware-based \sic{} countermeasures, a piece of code compiled for a generic \rv{} CPU in a cloud server cannot make safe assumptions about the microarchitecture on which it is running.
Existing tools for aiding software-level precautions by checking \sic{} vulnerabilities on source code or x86 binaries are not compatible with \rv{} machine code.

In this work, we study the requirements and goals of architecture-specific leakage analysis for \rv{} and illustrate how to achieve these %
goals with the help of fast and precise dynamic binary analysis.
We implement all necessary building blocks for finding side-channel leakages on \rv{}, while relying on existing mature solutions when possible.
Our leakage analysis builds
upon the modular \sic{} analysis framework \mw{}, that examines execution traces for leakage through secret-dependent memory accesses or branches.
To provide suitable traces, we port the ARM dynamic binary instrumentation tool \mambo{} to \rv{}.
Our port named \mambov{} can instrument arbitrary binaries which use the 64-bit general purpose instruction set.
We evaluate our toolchain on several cryptographic libraries with \rv{} support and identify multiple leakages.

\end{abstract}

\section{Introduction}

Executing workloads in cloud environments with shared hardware resources is becoming more and more important, promising great flexibility and scalability. From a security viewpoint, however, this trend comes with a number of challenges, as shown by manifold examples of attacks that exploit microarchitectural side-channels in cloud systems~\cite{DBLP:conf/ches/InciGIES16,DBLP:conf/uss/YaromF14,DBLP:conf/sp/IrazoquiES15}.

While most of these cloud systems and the corresponding attacks are based on the conventional x86 architecture, a new architecture called \rv{} is gaining traction in both embedded applications and general-purpose hardware.
The royalty-free license~\cite{asanovic2014instruction} of \rv{} enables affordable hardware through lower development costs, and helps innovation: For example, there now are several open-source CPU designs which can be analyzed and extended by anyone~\cite{DBLP:journals/tvlsi/StangherlinS22,DBLP:journals/jhss/KumarDGBHCM20,DBLP:conf/asiaccs/NasahlSWM21}, promising the development of new hardware features like secure trusted execution environments (TEEs) which avoid the issues of existing commercial solutions. The software support for the \rv{} platform is growing as well, with major compiler vendors adding backends for emitting \rv{} machine code, which in turn allows porting operating systems like Linux.

The growing importance of \rv{} in general-purpose and cloud computing, coupled with a wide spectrum of CPU designs from various vendors, still necessitates caution to prevent repeating the mistakes that caused a lot of security issues on the established platforms.
One particular example is \emph{microarchitectural timing leakage} in cryptographic libraries, where subtle differences in how the microarchitecture processes certain operations lead to exploitable leakages, allowing a co-located attacker running on the same hardware as the victim code to extract cryptographic secrets. 
By microarchitectural timing leakage, we refer to architectural traces only, excluding transient execution attacks.
As most of the existing \rv{} hardware finds usage in the IoT or the automotive domain, there has been more focus on physical attacks like power side-channels, and little work on analyzing the co-location scenario so far. However, it is likely that many attack vectors from x86 and ARM will apply to \rv{} systems as well.
While there are several proposals for hardware countermeasures that would address this issue (e.g., resistant cache designs~\cite{DBLP:conf/uss/WernerUG0GM19,DBLP:conf/uss/DessoukyFS20,DBLP:conf/uss/SaileshwarQ21}), it is unlikely that all CPU vendors will include one of those mitigations in their processors. Thus, absent a proven hardware-based countermeasure, software-level mitigations are needed.

By now, most established libraries address timing leakages by employing so-called \emph{constant-time code}, i.e., code that exhibits the same control flow and memory access pattern independent of its secret inputs. However, the new compiler backends and different instruction set of \rv{} may re-introduce leakage previously fixed at source level~\cite{DBLP:conf/uss/AlmeidaBBDE16,DBLP:journals/acm/DanielBR22}. In addition, there is ongoing work on assembly-level implementations of cryptographic primitives, which are carefully optimized to fully utilize the underlying hardware to achieve best performance~\cite{target-scl-metal}, but may have subtle leakages. While there are lots of approaches for finding leakages on source-level or via generic languages, those cannot detect leakage introduced by the compiler. Finally, most of the corresponding proof-of-concept implementations lack usability~\cite{DBLP:conf/sp/JancarFBSSBFA22} or do not apply to \rv{}.

In this work, we discuss the requirements of analyzing \rv{} software for side-channel leakages, and show how an established side-channel analysis framework can be adapted to also support \rv{} binaries. For that, we build upon the \emph{Microwalk} framework~\cite{DBLP:conf/ccs/WichelmannSP022}, that analyzes execution traces in order to identify vulnerabilities, and then yields a detailed leakage report. While Microwalk generates its execution traces through dynamic binary instrumentation (DBI), no such tool is yet available for \rv{}. Thus, we develop the first DBI tool for \rv{}, called \mambov{}, which sets up on the MAMBO toolkit~\cite{DBLP:journals/taco/GorgovanDL16} for ARM, and show how we can use this tool to generate Microwalk-compatible traces. We evaluate our leakage analysis toolchain on several cryptographic libraries with support for \rv{}, and uncover multiple vulnerabilities.

\subsection{Our Contribution}
In summary, our contributions are:
\begin{itemize}
    \item We analyze the similarities and differences between \rv{} and established architectures in terms of side-channel vulnerabilities, and extract requirements for building side-channel-resistant software on \rv{}. %
    \item We implement \mambov{}, a \rv{} port of the ARM-based DBI tool MAMBO, enabling us to natively instrument \rv{} binaries.
    \item We include \mambov{} in the \mw{} framework for finding timing side-channels in software binaries, building the first toolchain for automatically analyzing \rv{} programs.
    \item We analyze several \rv{} builds of cryptographic libraries and detect various leakages. %
\end{itemize}

The source code is available at \url{https://github.com/UzL-ITS/MAMBO-V}.

\subsubsection{Responsible Disclosure}
We disclosed the potentially exploitable AES vulnerabilities to the developers of the respective libraries, who all acknowledged our findings. They were mostly aware of the issues of the relevant implementations, and \wolfssl{} and \openssl{} have (undocumented) compiler flags which partially fix the leakages (see \cref{sec:evaluation:vulnerabilities}). At the time of submission, there is ongoing work on patches that ensure that the default implementations are secure, or on appropriate documentation changes.

\section{Background}

\subsection{\rv{}}
\rv{} is a reduced instruction set computer (RISC) load-store architecture, with a focus on broad availability through permissive licensing
and high modularity to support all applications from small low-power IoT devices over personal mobile devices to large-scale general purpose computers. Its open-source character allows easy extensibility through a so-called base-plus-extension instruction set architecture (ISA).  As a RISC architecture, only designated instructions operate on memory, whereas the arithmetic merely happens in registers.
The most important standardized extensions for \rv{} are I, M, A, C, F, D, Zicsr and Zifencei, which are often grouped together as \emph{RV64GC}. Also, more specialized extensions are drafted and partially ratified, such as the vector extension and scalar cryptographic extension~\cite{riscv-crypto-extensions}.
Instruction encodings are designed to simplify hardware implementations to increase performance and efficiency \cite{DBLP:phd/us/Waterman16}.%

\subsection{Dynamic Binary Instrumentation}
Binary instrumentation allows inserting code into an existing binary in order to monitor or modify the program's behavior. The insertion points are determined through user-supplied rules or callback functions.

\emph{Static binary instrumentation} (SBI), also called binary rewriting, permanently inserts instrumentation code into the binary in an offline phase~\cite{DBLP:conf/sp/DineshBXP20}.
While this approach promises a small runtime overhead, it is error-prone due to relying on a correct disassembly of the program. In addition, SBI cannot handle special cases like just-in-time compilation or self-modifying code. %

In \emph{dynamic binary instrumentation} (DBI), the instrumentation code is added with the help of an instrumentation framework at runtime.
The DBI framework combines application and instrumentation code and executes the resulting code directly on the target platform.
DBI engines introduce a slightly higher overhead than SBI due to the code translation at runtime, but most prevalent instrumentation frameworks feature optimizations like caching, so each code block needs to be instrumented only once.
Popular DBI engines include Intel Pin~\cite{DBLP:conf/pldi/LukCMPKLWRH05}, DynamoRIO~\cite{DBLP:conf/cgo/BrueningGA03}, QBDI~\cite{qbdi2017quarkslab} and the heavyweight analysis framework Valgrind~\cite{DBLP:journals/entcs/NethercoteS03}, which were initially built for x86 and then, in some cases, extended to also support other architectures like ARM's \arm{} and \aarch{}. 

However, as ARM is a RISC architecture and thus quite different to x86, x86-specific optimizations in a DBI engine may have little or even negative effects. %
\mambo{}~\cite{DBLP:journals/taco/GorgovanDL16} is a DBI tool specifically designed and developed for ARM, making it suitable for efficiently handling RISC architectures.
In addition to some ARM-specific optimizations, \mambo{} has general DBI features like a cache for storing already instrumented code and scanning new code in basic block units.
Moreover, it supports behavioral transparency, which means that the execution of all ABI-compliant binaries is guaranteed to be correct. The application binary interface (ABI) defines the calling convention, which includes register allocation for parameters and stack pointer behavior.

\subsection{Microarchitectural Side-Channels}\label{sec:bg:uarch-scs}
In a cloud setting, usually, many processes from different customers share the same underlying hardware.
These processes may work with sensitive data, which should not be leaked to an attacker. While there are many architectural safeguards in place to prevent data from flowing from one process to another directly, there are more subtle \emph{side-channels} that use properties of the underlying microarchitecture to extract some information from the running code. One prominent example are so-called \emph{cache attacks}~\cite{bernstein2005cache,DBLP:conf/ctrsa/OsvikST06,DBLP:conf/ches/AciicmezBG10,DBLP:conf/uss/YaromF14}, where the attacker brings the (shared) CPU cache into a known state, and then monitors changes to this state in order to learn whether the victim has accessed data within a certain address range. This way, the attacker can infer the code line the victim is currently executing, or determine the index of a table lookup. Besides the cache, there are many more shared resources that the attacker can monitor and exploit, like the translation look-aside buffer~\cite{DBLP:conf/uss/GrasRBG18} and the branch prediction unit~\cite{DBLP:conf/ctrsa/AciicmezKS07}. Note that we only consider attacks that target architectural traces, so transient execution attacks like Spectre~\cite{DBLP:conf/sp/KocherHFGGHHLM019} are out-of-scope.

A commonly used software-based countermeasure against \sic{} attacks is \ct{} code without any secret-dependent memory accesses or branches~\cite{DBLP:conf/uss/AlmeidaBBDE16}. This code exhibits the same control flow and data flow independent of the processed secret, so a side-channel attacker cannot learn anything by looking at an execution trace as provided by a cache attack. As cryptographic implementations are a primary target for side-channel attacks, most current cryptographic libraries feature constant-time code.

\subsubsection{Leakage Detection Tools}
To ease checking implementations for side-channel vulnerabilities, numerous tools and approaches have been proposed.
Tools that analyze source code include \emph{ct-fuzz}~\cite{DBLP:conf/icst/HeEC20} that uses a specialized form of fuzzing, \emph{ct-verif}~\cite{DBLP:conf/uss/AlmeidaBBDE16} based on formal verification methods and \emph{CaSym}~\cite{DBLP:conf/sp/BrotzmanLZTK19} that symbolically executes the source code.
Moreover, there are various tools that analyze binaries through static techniques, like \emph{BINSEC/REL}~\cite{DBLP:journals/acm/DanielBR22} using symbolic execution, \emph{CacheS}~\cite{DBLP:conf/uss/0011BL0ZW19}
combining symbolic execution with taint analysis, or \emph{CacheAudit}~\cite{DBLP:conf/uss/DoychevFKMR13} which uses formal methods to find leakages on all paths of a program.
Finally, dynamic binary approaches comprise statistical timing measurements like in \emph{dudect}~\cite{DBLP:conf/date/ReparazBV17}, constraint modeling in \emph{Abacus}~\cite{DBLP:conf/icse/BaoWLLW21}, as well as trace alignment in \emph{DATA}~\cite{weiser2018data} or trace merging in \mw{}~\cite{DBLP:conf/ccs/WichelmannSP022}.

\section{Overview}
\label{sec:overview}

\begin{figure}[t]
    \centering
    \includegraphics[width=0.7\textwidth]{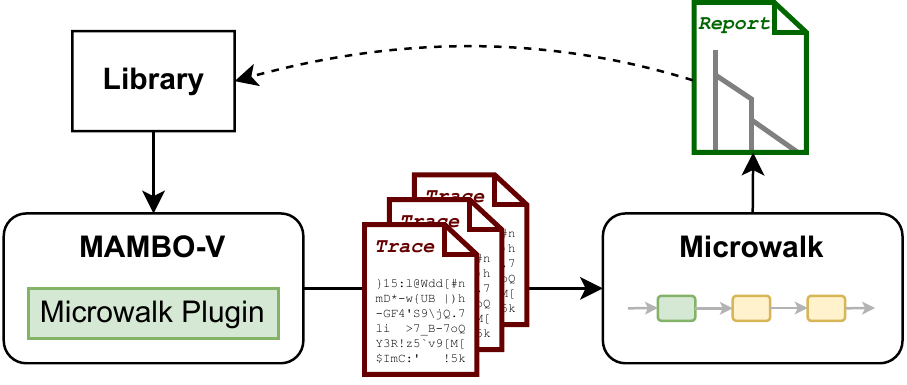}
    \caption{\rv{} side-channel analysis overview. \mambov{} instruments a \rv{} library and generates execution traces, which are subsequently analyzed using Microwalk. The resulting analysis report then helps the developer to find and fix the identified leakages.}
    \label{fig:overview}
\end{figure}

We first describe requirements and our approach for analyzing the side-channel security of \rv{} implementations running in a co-located setting.

\subsection{Analysis Approach}\label{sec:overview:approach}
As described in \Cref{sec:bg:uarch-scs}, there are numerous tools and approaches for finding side-channel leakages in software. Any useful tool should unify the following properties~\cite{DBLP:conf/sp/JancarFBSSBFA22,DBLP:conf/ccs/WichelmannSP022}: First, it should accurately localize the respective leakages, so the developer can directly understand the cause of a leakage and start building a patch. Then, the analysis should be fast enough, so there is immediate feedback whenever there is a code change.
Finally, to aid adoption in the developer community, the tool should not be too hard to set up and use.%

To check whether \rv{} code is leakage-free, focusing on the source code alone is insufficient.
For example, there have been cases where a misguided compiler pass ``optimized'' constant-time code, producing binaries with leakages that are not present in the source code~\cite{DBLP:conf/cans/KaufmannPVV16,DBLP:conf/uss/AlmeidaBBDE16}.
Daniel et al.~\cite{DBLP:journals/acm/DanielBR22} further provide an extensive evaluation of different compiler versions, optimization levels and target architectures, showing that \ct{} properties always need to be validated on the binary level.
Compiling the code for x86 and using existing analysis tools is not sufficient either, as x86 compilers may use different optimization passes than \rv{} compilers.
In addition, x86 has special extensions like AES-NI or the \texttt{pclmulqdq} instruction for carry-less multiplication (used in Galois counter mode), which may substitute otherwise leaking code paths.

The necessity to work with \rv{} specific assembly leaves the option to use either static or dynamic binary analysis.
While static binary approaches offer some guarantees that purely dynamic tools cannot give, they often suffer from poor performance and require lots of manual interaction. On the other hand, dynamic analysis is heavily dependent on the achieved coverage, i.e., leakage can only be found in code that is actually executed. However, for cryptographic implementations, it was found that a small number of random test cases is sufficient to cover the relevant code~\cite{weiser2018data,DBLP:conf/ccs/WichelmannSP022}. In addition, dynamic analysis is easy to use, as the user only has to call the respective primitives.

\subsection{Toolchain}
With the aforementioned requirements in mind, we picked the \mw{} framework~\cite{DBLP:conf/acsac/WichelmannMES18,DBLP:conf/ccs/WichelmannSP022} as a basis for our \rv{} leakage analysis. \mw{} uses DBI to generate execution traces from user-supplied programs, and offers several analysis modules that compare these traces in order to find leakage. While the authors originally designed \mw{} for x86 binaries, its modular structure and generic trace format encourage addition of trace generators for other architectures.

This leaves the problem of generating Microwalk-compatible execution traces for \rv{}.
At the time of writing, there is no generic DBI framework for \rv{} available, that offers the necessary flexibility for generating the information \mw{} needs.
Another requirement is transparency, such that the execution traces are not influenced by the DBI engine itself, which would otherwise distort the analysis result.
Instead of building a new DBI framework, we decided to port an existing framework for another RISC architecture, that is MAMBO~\cite{DBLP:journals/taco/GorgovanDL16} for ARM. The similarities between ARM and \rv{} allow us to reuse most of the general-purpose logic from MAMBO, like plugin handling or memory management. Our port, named \mambov{}, implements the most significant performance optimizations from \mambo{}, which are inline hash table lookups and direct branch linking. Additionally, we add support for atomic sequences, which need special handling on \rv{} hardware. %
We are working with the maintainers of MAMBO to contribute our \rv{} patches to the main project.

The resulting toolchain is illustrated in \cref{fig:overview}.

\section{\mambov{} Implementation}

We now describe our \rv{} port of the MAMBO DBI framework, named \mambov{}. We give an overview over its generic features and discuss notable performance optimizations as well as \rv{} specifics to be considered.

\subsection{Instrumentation Approach}
\subsubsection{Target Platform}
\mambov{} targets RV64GC platforms, i.e., processors with support for the RV64I base instruction set and its most common extensions. %
Like MAMBO, \mambov{} aims for \emph{behavioral transparency}: Binaries that are compliant to the standard \rv{} ABI are executed correctly. %
This does not affect the correctness %
of our side-channel analysis, as we can expect that compilers emit standard-compliant code and that the analyzed programs are not malicious.

\subsubsection{Execution Model}
Just as the ARM implementation of \mambo{}, \mambov{} unifies the instrumentation framework and the target application in a single process.
On startup, a custom ELF loader reads the \rv{} ELF file and potential dependencies of the target application into the memory of the \mambov{} process, such that the engine can access the target's full code. After initialization is done, \mambov{}'s dispatcher proceeds loading and translating chunks of the target's code on-the-fly, while inserting instrumentation at the points specified by the user. Each chunk consists of a single basic block, i.e., a sequence of instructions with a single entry point at the beginning and a single exit point at the end. This way, the dispatcher can safely hand over control to the translated chunk, and reclaim it after the chunk has fully executed.

\subsubsection{Plugin API}
In order to facilitate the usage of \mambov{} for application developers who want to analyze their applications, we also ported the plugin API from \mambo{}. A plugin contains user-supplied functions, which are called at certain events, e.g., when translating a basic block. With these functions, the user can then insert instrumentation code during translation. Other supported events are function entry/exit, threads and system calls. In our analysis, we primarily utilize the instrumentation to insert trace writing code.

\subsubsection{Optimizations}
To speed up analysis, we have ported a number of performance optimizations from \mambo{}. Most of the overhead that arises during DBI comes from the code translation and context switches between the dispatcher and the target application.
The most notable optimization is the code cache, which is a common feature of DBI frameworks: It is located outside the target application's address space and stores a limited amount of translated basic blocks. This avoids re-translation of frequently executed code, improving overall performance significantly. Other optimizations are hash tables %
for faster resolution of translated blocks and direct branch linking to speed up jumping between different blocks in the code cache without invoking a costly context switch to the dispatcher.

\subsection{New Features for \rv{}}

\subsubsection{Atomic Sequences}
A challenge we encountered on \rv{} cores are tightly constrained \emph{atomic sequences}, which ensure exclusive memory operations for multiprocessor systems and process synchronization.
Software locks for resources that should only be accessed by a single thread or process at a time are often translated to \emph{atomic loops} by the compiler. An atomic loop contains an atomic sequence, which begins with a load-reserved (\texttt{LR}) instruction and ends with a store-conditional (\texttt{SC}) instruction. The atomic loop loops over the atomic sequence until the \texttt{SC} eventually succeeds.
The result of the \texttt{SC} instruction depends on whether the reserved value was accessed during the atomic sequence and on the environmental constraints defined by the ISA. Among others, the ISA defines
 a maximum of 16 consecutive instructions between \texttt{LR} and \texttt{SC}, and allows only the base (I) instruction set, disallowing loads, stores, backward jumps or calls.

\begin{figure}
\begin{subfigure}[t]{.38\textwidth}
    \centering
    \includegraphics[]{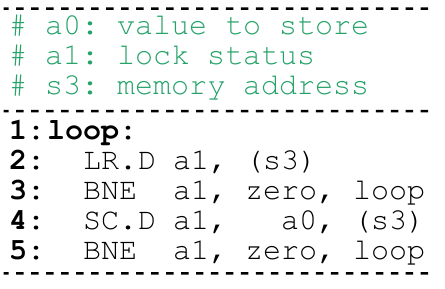}
    \caption{Original lock-acquire-loop.}
    \label{fig:RV-lock-acquire-loop}
\end{subfigure}
\begin{subfigure}[t]{.57\textwidth}
    \centering
    \includegraphics[]{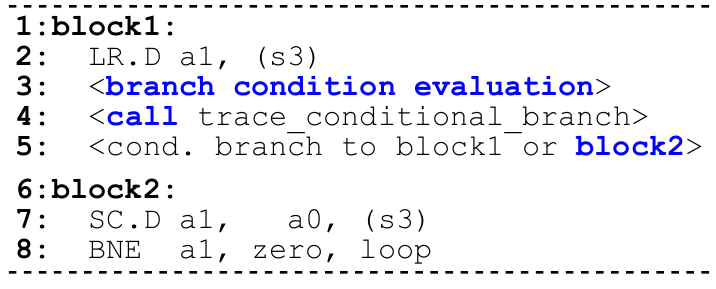}
    \caption{Instrumented lock-acquire-loop.}%
    \label{fig:RV-lock-acquire-loop-instrumented}
\end{subfigure}
\caption{Exemplary instrumentation of a lock-acquire-loop: The instrumentation may insert unconstrained instructions (marked in \textbf{\textcolor{blue}{blue}}) into the atomic sequence, e.g., add a function call with parameters to trace a conditional branch instruction. In order to set the argument registers, the original register contents have to be written to the stack using an unconstrained store instruction.}
\label{fig:LRSC-instrumentation-example}
\end{figure}

While the compiler enforces the constraints within an atomic sequence, the instrumentation done by \mambov{} can insert arbitrary instructions that break one of the above constraints. 
\Cref{fig:LRSC-instrumentation-example} shows an example of how
a direct port of \mambo{} would add unconstrained instructions to an atomic sequence:
First, the original loop in \cref{fig:RV-lock-acquire-loop} is split into two blocks
because of the conditional branch in line 3. 
Then, the resulting code cache blocks undergo optimization
and are instrumented as shown in \cref{fig:RV-lock-acquire-loop-instrumented}, leading to the insertion of unconstrained instructions 
(line 3-5).
The result is a non-sequential sequence that includes loads, stores, calls, and potential backward jumps, and is therefore not guaranteed to succeed on RISC-V.
However, requiring all instrumentation to adhere to the constraints would cause some instrumentation features to be lost in the process. %

On ARM, where atomic sequences are available as well, \mambo{} allows users to freely insert instrumentation, which when breaking a constraint causes undefined behavior, but does not affect stability on ARM Cortex processors. However, on our SiFive U54 core, violating a constraint can block the \texttt{SC} instruction from succeeding entirely, leaving the process stuck in a deadlock.
We encountered such a deadlock when instrumenting the dynamic linker. %

Thus, for reliable instrumentation on \rv{} cores, we designed a lightweight and behaviorally transparent solution for handling atomic sequences:
We use hardware-assisted software emulation to relax the hardware constraints by replacing the \texttt{LR} and the \texttt{SC} instructions. 
The \texttt{LR} is replaced by an equivalent normal load instruction, which marks the beginning of the software-emulated atomic sequence. 
To emulate the reserve, we also back up the original value for later comparison. 
The subsequent code is not bound by constraints anymore and safe for arbitrary instrumentation. 
Finally, we replace the \texttt{SC} instruction with a semantically equivalent atomic sequence that conditionally stores the new value if the value at the destination is equal to the previously created backup. 
Since we include a native atomic sequence to check for changes at the destination, our emulation remains thread-safe.
The observable behavior of the emulated atomic sequence is nearly identical to the original, with the only difference being that the emulation cannot detect stores on the reserved value that do not modify it. To the best of our knowledge, this difference does not effectively change the semantics of the emulated sequence, and therefore the traces remain identical.

\subsubsection{Global Pointer and Thread Pointer Register}
In contrast to ARM, the \rv{} standard calling convention defines a global pointer register \texttt{gp} and a thread pointer register \texttt{tp}. Applications use these registers to access structures such as the global offset table and global/thread-local variables. \mambov{} does not share these structures with its client, so \texttt{gp} and \texttt{tp} must be updated on each of the context switch between \mambov{} and the client. Originally, on ARM, a unidirectional context switch was sufficient, as the dispatcher does not make assumptions on register contents on entry. Thus, only the context of the client is fully saved when entering the \mambov{} context and restored when leaving again.
To support the distinct \texttt{gp}/\texttt{tp} contexts on \rv{}, we implemented a full context switch for these two registers, while keeping the unidirectional context for all other registers to minimize the overhead.

\subsubsection{Shorter Jump Encoding}
\rv{} and ARM do not have direct branch instructions that take an absolute immediate address. Due to different instruction encodings, the maximum range of ARM branch instructions is $\pm$128 MiB, while on \rv{} it is only $\pm$1 MiB. The code cache in \mambov{} can be much larger than 1 MiB. Hence, for \mambov{}, we decided to use indirect jumps to transfer control flow back to the dispatcher. Loading the address and performing the jump takes 14 additional bytes in the code cache, but due to the long lifetime of translated code and runtime overhead of the client-dispatcher context switch the effect on the overall performance and memory consumption is negligible.

\section{Side-Channel Leakage Analysis}
In the following, we describe our approach for finding architecture-specific leakage in code compiled for \rv{} with the help of \mambov{}.
We focus on implementations of cryptographic algorithms, as their impact on the security of systems and communication is high. However, the concepts do apply to any scenario where secret information should not be exposed to an attacker recording execution traces.
As discussed in \cref{sec:overview}, source-level analysis is often not sufficient, and binaries may contain leakages even though the original source code is constant-time. Therefore, we opted for a binary approach based on \rv{}-specific DBI for execution trace generation and \mw{} for leakage analysis.

\subsection{Leakage Model}
We adopt the leakage model as specified for \mw{}~\cite{DBLP:conf/ccs/WichelmannSP022}: We supply the attacker with an implementation, a number of secret inputs and corresponding \emph{execution traces}. An execution trace consists of a sequence of all executed instructions and accessed memory addresses, but does not contain actual processed data. The attacker also gets access to all public inputs and outputs. We consider the implementation constant-time if all traces are identical, i.e., when the attacker does not learn anything about the secret input by looking at a trace. In other words, in a constant-time program, the observed control flow and memory accesses are independent of the secret inputs.

This leakage model assumes a rather strong attacker, as the known side-channel attacks can only retrieve a fraction of the information expressed in a full execution trace. For example, cache attacks are limited to granularities of 32 or 64 bytes on most systems, and control flow tracking techniques like single-stepping only work in very specific scenarios. Due to the lack of suitable hardware, there has not yet been much work on side-channels for \rv{}. Thus, while we expect similar vulnerabilities on upcoming \rv{} processors as are already known for other architectures, sticking to a strong leakage model is the safest way forward.
We only consider secret-dependent control flow and memory accesses that are architecturally reachable, so transient execution attacks are out-of-scope.

\subsubsection{Implementation in \mw{}}
\mw{} implements the above leakage model through a simple dynamic analysis pipeline, which generates secret inputs (called \emph{test cases}), collects and preprocesses corresponding execution traces, and finally compares those traces with each other. If \mw{} finds a difference between two or more traces at a given code position, this difference is reported as \emph{leakage}, as an attacker may exploit this difference to tell apart two or more secret inputs. If all traces are identical, the attacker does not learn anything about the underlying secret inputs, and the implementation is reported as non-leaking.

\subsection{Required Information}
\begin{figure}[t]
    \centering
    \includegraphics[width=\textwidth]{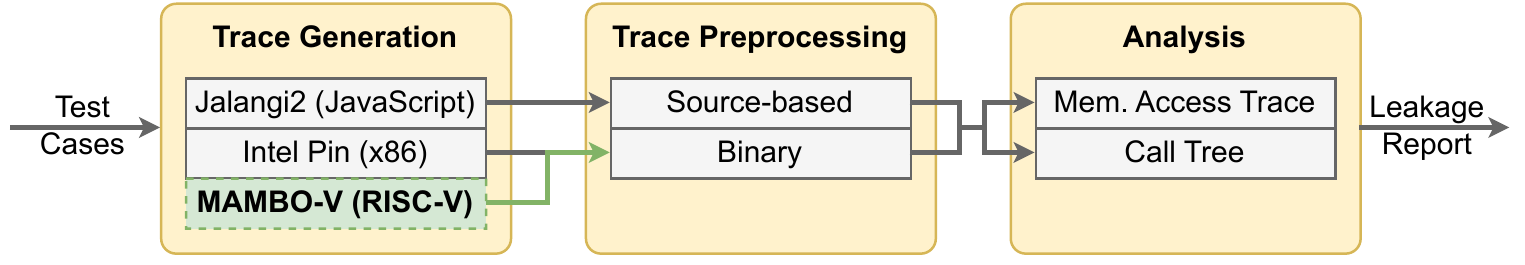}
    \caption{\mw{} pipeline with a new trace generation module based on \mambov{}. Each trace generation module may emit either source-based or binary execution traces, which are then preprocessed into a common trace format that can be parsed by all analysis modules.}
    \label{fig:microwalk-with-mambo}
\end{figure}

\mw{} uses a common generic execution trace format to run its analysis modules on, so we build a toolchain that collects \rv{} execution traces and converts them into \mw{}'s format. \mw{} already offers two raw trace preprocessors, one for converting source-based execution traces from languages like JavaScript, and another one for binary traces from compiled code. While the binary trace preprocessor was originally written for x86, we found that its raw trace format is generic enough to also be used on other architectures. We thus only need to create a trace generator for \rv{}, that emits raw execution traces in the same format as the existing Intel Pin module (\cref{fig:microwalk-with-mambo}).

A raw binary execution trace from \mw{}'s Intel Pin module combines the following information:
\begin{itemize}
    \item taken/non-taken branches, with source and (if applicable) target address;
    \item memory accesses, with instruction address and accessed memory address;
    \item heap/stack allocation blocks, with start and end address;
    \item start and end addresses of the memory-mapped executable binaries.
\end{itemize}
We collect this data using a plugin for the \mambov{} DBI framework.

\subsection{\mambov{} Trace Plugin}
\label{sec:trace_plugin}

\subsubsection{Interaction with the Target Program}
In order to analyze a cryptographic primitive, the primitive has to be made available to the DBI framework. We follow \mw{}'s approach by asking the user to supply a small function that receives a test case file with secret inputs and then calls the cryptographic primitive. Our \mambov{} plugin registers a \emph{function call} event callback for detecting execution of that function, so it can detect when test case execution starts and ends. This method has the advantage that we do not need to re-instrument the binary for each test case, but can reuse the existing instrumentation, which speeds up trace generation significantly. Before the first test case begins, we record a \emph{trace prefix}, that contains initializations of all global objects that may be referenced during test case execution.

\subsubsection{Recording Control Flow and Memory Accesses}
When a test case begins, which is signaled by the respective event callback, our plugin opens a new binary trace file. We also register an instrumentation callback, which is called whenever a new basic block is instrumented. %
In this callback, we check each instruction for control flow and memory accesses, and add instrumentation to that instruction if necessary. %
The resulting instrumented code then writes to the trace file whenever the respective instruction is executed. %
To avoid tracing information outside our target functions, the plugin receives a list of binaries that should be traced. 

\subsubsection{Tracking Memory Allocations}
\mw{} needs both a list of allocated heap memory blocks and the regions of the memory-mapped executables. To collect heap blocks, we register \emph{function call} and \emph{function return} event callbacks for the \texttt{malloc}, \texttt{calloc}, \texttt{realloc} and \texttt{free} functions, and log their parameters and return addresses. For the static memory regions, we hook into the \emph{VM operation} event handler and extract the required information from \texttt{VM\_MAP} events, which are triggered whenever a new ELF file is loaded. %

\section{Evaluation}
To evaluate the performance of our toolchain and assess the current state of \sic{} security on \rv{}, we analyze a number of frequently used cipher and signature functions for several popular libraries.
We describe the experimental setup, analyze the performance of trace creation and analysis, and discuss and evaluate the discovered leakages. The results are summarized in~\Cref{tab:rv-mw-libs}.

\input{eval-results-rv.tex}

\subsection{Experimental Setup}

As described in \Cref{sec:overview}, we combine \mambov{} with \mw{} to natively analyze the leakage of binaries on \rv{}. 
We record the traces with \mambov{} on a Microchip PolarFire SoC FPGA Icicle Kit with four SiFive U54 cores featuring RV64GC.
The trace analysis with \mw{} is executed on an AMD Ryzen 9 7950X with 16 cores.

\subsubsection{Libraries}
Due to its modular structure, the \rv{} architecture allows for a broad range of target applications, from small embedded devices to
server CPUs.
To reflect this, we chose to analyze \wolfssl{}~\cite{target-wolfssl} and \mbedtls{}~\cite{target-mbedtls} as examples for libraries that support many architectures and that are optimized for the embedded market.
\openssl{}~\cite{target-openssl} and \nettle{}~\cite{target-nettle}, on the other hand, are general purpose cryptography libraries that are used across different architectures and chip sizes. 
In addition, as an example of a library specifically written for \rv{}, we investigated \scl{} (SiFive Cryptographic Library)~\cite{target-scl-metal}. Finally, as a reference for \ct{} implementations, we included \libsodium{}~\cite{target-libsodium}.

\subsubsection{Analyzed Primitives}
We wrote analysis wrappers for \aesecb{}, the authenticated encryption schemes \aesgcm{} and \chacha{}, and  the signature algorithms \edsig{} and \ecdsa{} (curve \texttt{secp192r1}; \texttt{secp256r1} for \scl{}). The wrappers
initialize the necessary environment and call the target functions, if supported by the respective library. We skipped the \ecdsa{} implementations in \nettle{} and \mbedtls{}, as those are comparably slow and thus lead to traces which exceed the limited resources of our evaluation platform. 

All libraries and target wrappers were cross-compiled with the \rv{} GNU Compiler Toolchain 12.2.0~\cite{riscv-gnu-toolchain} for RV64GC and ISA specification 2.2.
We built all libraries with default options and appropriate additional security flags as stated in their documentation. %
All libraries except \openssl{} are built with optimization level \texttt{-O2}.
\openssl{} was built with optimization level \texttt{-O3}.

\subsubsection{Test cases}
We generated 16 test cases for each primitive by creating 16 random keys, and supplied these test cases to the target function.
Since \mw{} measures differences in the execution traces, any other input outside the test cases must be kept constant to avoid false positives. %
Therefore, inputs such as %
initialization vectors were set to fixed values. Random values like the ephemeral key in \ecdsa{} were generated by custom test case-dependent RNGs.
We opted for using smaller key sizes, as the cryptographic procedures are invariant of the key size, and larger key sizes increase the resource consumption of the leakage analysis without uncovering further vulnerabilities~\cite{DBLP:conf/ccs/WichelmannSP022}.

\subsection{Performance Results}

The performance of the \sic{} analysis on \rv{} depends on the time required for tracing the target function and analyzing the traces. 
The runtime for all targets is summarized in~\Cref{tab:rv-mw-libs}.

\subsubsection{Tracing}
The duration of tracing 16 executions for each target is inherently constrained by the limited performance of the SiFive U54 core.
For the symmetric ciphers and \edsig{}, the tracing took at most a few minutes, which suggests that our toolchain is suitable for everyday use on a developer's computer. With newer and more performant \rv{} cores, the tracing time should further decrease.

One outlier is \openssl{}, where a majority of the tracing time was spent in the library initialization, which is mostly irrelevant for the leakage analysis. To reduce this overhead, the developer could disable most features when compiling the library for vulnerability evaluation and target low-level functions.

\subsubsection{Analysis}
With one exception, the trace preprocessing and analysis of nearly all targets took less than 5 seconds.
The fast analysis allows for frequent execution of any test. The outlier, \ecdsa{} for \openssl{}, was slowed down by preprocessing the huge traces, so optimizing the tracing time should fix this as well.

\subsection{Vulnerabilities}\label{sec:evaluation:vulnerabilities}

The leakage analysis for the chosen popular libraries shows many vulnerabilities across the board, except for \libsodium{} which only implements a limited number of ciphers and signature algorithms that allow for an implementation with better resistance against timing attacks by design. 
Indeed, all analyzed implementations of \chacha{} and \edsig{} are \ct{}. 
We summarize the results in~\Cref{tab:rv-mw-libs} in the columns ``\# Lkgs.'' (total leakages) and ``\# Uniq.'' (unique leakages).
An instruction or function can be called or reached from multiple contexts, thus potentially leaking different secrets with varying leakage severity. Therefore, we also count unique occurrences of leaking instructions.

In-depth analysis of the libraries %
showed that most 
provide specific assembly implementations for x86 and other architectures that use \ct{} primitives. 
For \rv{} though, due to lack of specifically optimized implementations, the libraries fell back to default ones, which often turned out to be non-constant-time, even when using the hardening flags specified in the documentation.

\subsubsection{Symmetric Ciphers}
All analyzed \aesecb{} implementations leak secret information through their timing behavior. 
The examined libraries do not provide \rv{}-specific code, but fall back to their default \clanguage{}/\cpplanguage{} implementations, which use
either T-table or S-box lookups for AES encryption and round key generation. Previous work has shown that table lookups are exploitable by timing measurements~\cite{bernstein2005cache}.
The number of unique leakages varies between the different libraries depending on whether the encryption rounds are unrolled and how the final step is scheduled. After informing the \openssl{} developers that we found several leakages in the default \aesecb{} implementation, we were pointed to an undocumented compiler flag that enables an alternative AES implementation, which we verified to be constant-time. However, they also stated that the flag leads to a 95\% performance loss, which is why it is not enabled by default.

The authenticated encryption algorithm \aesgcm{} builds upon the same primitives as \aesecb{} and thus also shows the same table lookup leakage for the encryption step.
In addition, the GCM mode adds authentication through computation of a GHASH, which involves encryption of a 128-bit string of zeros and the IV.
The result of the latter encryption is used for the final computation of the authentication data. %
The multiplication used for the GHASH is implemented with a hash lookup table, where the accessed index depends on the current ciphertext and the hash value of the previous block.

We compared the leakage result of \aesgcm{} on \rv{} for the libraries \openssl{} and \mbedtls{} against the analysis on x86. While the \rv{} binaries contain many leakages as explained above, we observed no leakages for x86 binaries. 
The x86 implementations use the AES-NI hardware extension for encryption and the \texttt{clmul} extension for computation of the GHASH.
Until such extensions are available for \rv{}, cryptographic libraries must feature \ct{} software implementations. For \wolfssl{}, we learned during disclosure that there is a \texttt{GCM\_SMALL} flag, which enables a non-table-based GHASH implementation. While designed (and documented) primarily for small code size, we found that it is constant-time and thus a secure alternative for the default implementation.

\subsubsection{Asymmetric Signature Algorithms}
None of the analyzed implementations of \edsig{} shows any \nct{} behavior, emphasizing its inherent resistance against timing attacks, even though there are no specific assembly implementations for \rv{}. 
However, we found leakage for all analyzed implementations of \ecdsa{}, especially in the implementation from \openssl{}. 
Even the specially crafted \rv{} implementation from \scl{} reveals \nct{} behavior, though the library is not yet deemed production-ready.
Despite the high number of potential vulnerabilities, we found that all analyzed \ecdsa{} implementations use blinding, rendering the discovered leakages likely unexploitable.

\section{Discussion and Future Work}

\subsubsection{Limitations of \mw{}}
As we base our analysis on \mw{}, we inherit some of its limitations.
Currently, \mw{} only supports deterministic implementations. Thus, all entropy must come from the secret inputs. While this scenario works well with symmetric and constant-time asymmetric cryptographic primitives, it has some issues with blinded implementations which obscure the computation by randomizing the input parameters. Disabling the randomness is not sufficient either, as this would just expose leakages which are normally obscured by blinding. As a solution, \mw{} should be extended to support randomized implementations. Another limitation of \mw{}'s analysis algorithm is the possibility of several small leakages higher up in the call chain hiding leakages further down, though we did not observe this during our evaluation. Finally, %
\mw{}'s dynamic approach heavily depends on the coverage. While it was found that few random test cases usually suffice~\cite{weiser2018data,DBLP:conf/ccs/WichelmannSP022}, the user should check that all relevant code locations have been reached.

\subsubsection{Other Applications of \mambov{}}
While we used \mambov{} for generating execution traces, the tool is far more versatile.
The plugin API supports a variety of different callbacks, making it on par with other widely-used frameworks like Intel Pin. For example, new plugins can aid with control-flow checks or help in bug detection. The broad similarities to ARM allow reusing analysis code originally written for MAMBO with little adjustments.

\subsubsection{Leakage Analysis on ARM}
The proximity of \rv{} and ARM suggests that the \mambov{} trace generator plugin can be ported to the original MAM\-BO implementation with little adjustments. With that plugin, one could generate execution traces from ARM binaries, and analyze these traces for side-channel vulnerabilities using \mw{}, yielding a dynamic leakage analysis toolchain for ARM. Thus, our toolchain comprising a tracer plugin and \mw{} provides a solid basis for fast and accurate side-channel leakage analysis on various systems.

\section{Related Work}

\subsubsection{Analysis of Code on Intermediate Representations} Instead of instrumenting code natively, the machine code can be lifted to a generic intermediate representation. This approach is taken by the ongoing \rv{} port~\cite{valgrind23riscvport} of the heavyweight instrumentation framework Valgrind~\cite{DBLP:journals/entcs/NethercoteS03} and the full-system emulator QEMU~\cite{DBLP:conf/usenix/Bellard05}, which do an emulated analysis of \rv{} instructions on the intermediate representations of the respective framework.
Thereby, it is possible to re-use existing analysis tools like memory leaks detection or call graphs.
Apart from that, the whole system reverse engineering tool PANDA~\cite{DBLP:conf/acsac/Dolan-GavittHHL15} provides a way to capture an execution trace, replay it afterwards and combine it with extensive analysis through different plugins.
However, emulated analysis meets a different objective than analyzing architecture-specific leakage, as the leakage may be hidden during lifting to the intermediate representation. 
Furthermore, the emulators impose a very high overhead and are too resource-consuming to use them in restricted environments or for an efficient analysis with \mw{}.

\subsubsection{Side-Channel Analysis} Side-channel attacks on \rv{} are receiving growing attention by security research.
Apart from the timing \scs{} we analyze in this work, there have been efforts to secure \rv{} implementations against leakage through power \scs{}~\cite{DBLP:conf/dac/MulderGH19}. Further, electromagnetic leakage builds the basis for a successful fault attack in~\cite{DBLP:journals/tches/NashimotoSUH22}, showing that manifold leakage channels need to be addressed.
As some \rv{} systems also support out-of-order execution, they are susceptible to Spectre~\cite{DBLP:conf/sp/KocherHFGGHHLM019} attacks~\cite{gonzalez2019replicating,le2020experiment}. Recently, it was shown that data can be leaked from speculative execution through cache attacks~\cite{le2023cross}.
The vulnerability to Spectre-style attacks further motivates the development of a framework to automatically detect timing \scs{} in software, %
because apart from direct exploitation, the timing differences can also be used as a way to leak speculatively accessed secrets.

\subsubsection{Hardware-Based Countermeasures}
A \rv{} working group developed a number of extensions intended for secure cryptography, which were ratified in 2022~\cite{riscv-crypto-extensions}. This includes hardware-acceleration for symmetric encryption and hash functions, but also the \emph{Zkt} extension, which specifies constant-time properties for certain instructions. If a vendor implements the Zkt extension, certain arithmetic instructions are guaranteed to have data-independent execution time.
However, solely instruction-based approaches are insufficient, as most vulnerabilities are caused by higher-level data-dependent behavior.
Yu et al. propose support for
oblivious memory accesses, which would block most timing \scs{}~\cite{DBLP:conf/ndss/YuHHF19} and thus go far beyond simply avoiding data-dependent instruction latency like in the Zkt extension.
With hardware-integrated fully automated Boolean masking~\cite{DBLP:journals/tvlsi/StangherlinS22}, hardly any software-level precautions need to be taken against power \scs{}.
To protect against data leakages in ALU, memory and memory interfaces, INVITED~\cite{DBLP:conf/dac/MulderGH19} uses state-of-the-art masking techniques.

However, these hardware mechanisms are always applied, not only for secret inputs, making the solutions potentially inefficient for workloads where only a small fraction of all executed instructions is truly security-critical.
Moreover, in a cloud scenario, the clients have limited control about the hardware actually used, making secure software implementations indispensable.

\section{Conclusion}
In this paper, we have presented the first comprehensive \sic{} analysis for implementations of cryptographic primitives on \rv{}.
We have shown that
some of the most popular open-source cryptographic libraries lack proper \sic{} resistance on \rv{}.
For our work, we have studied the requirements for leakage detection on \rv{} and designed a thorough approach to incorporate all requirements into a mature \sic{} analysis framework that we have extended with all necessary building blocks.
We have based our analysis toolchain on \mw{} and augmented the framework with the necessary \rv{} specific tracing capabilities by implementing the DBI tool \mambov{}.
Our evaluation pinpoints several potentially exploitable leakages that should be fixed by the developers and emphasizes the need for complete and precise \sic{} analysis capabilities on \rv{} to pave the way for secure computations on shared \rv{} hardware in the cloud.

\subsubsection{Acknowledgements.} %
We thank the library maintainers for the smooth disclosure process, and the reviewers and our shepherd for their helpful comments and suggestions.
This work has been supported by DFG under grants 427774779 and 439797619, and by BMBF through projects ENCOPIA and PeT-HMR.

\bibliographystyle{splncs04}
\bibliography{ref}

\end{document}

%% file: eval-results-rv.tex
\begin{table}[h!]
    \caption{Result of leakage analysis of several cryptographic libraries on \rv{}. ``Tr. CPU'' shows the CPU time for generating the raw traces and ``An. CPU'' the CPU time for trace preprocessing and analysis. The columns ``\# Lkgs.'' and ``\# Uniq.'' show the total and unique number of detected leaking code lines.}
    
    \label{tab:rv-mw-libs}
    \centering
    \begin{tabular}{ p{38mm} p{18mm} >{\raggedleft\arraybackslash}p{15mm} >{\raggedleft\arraybackslash}p{15mm} >{\raggedleft\arraybackslash}p{15mm} >{\raggedleft\arraybackslash}p{15mm} }
        Target & Type & Tr. CPU & An. CPU & \# Lkgs. & \# Uniq. \\ \hline \hline

        \multicolumn{4}{l}{\textbf{\wolfssl{}}~\cite{target-wolfssl} 5.5.4} \\
            \quad \aesecb{} & cipher & 1 sec & $<$ 1 sec & 157 & 157 \\
            \quad \aesgcm{} & aead-cipher & 2 sec & $<$ 1 sec & 493 & 184 \\
            \quad \chacha{} & aead-cipher & $<$ 1 sec & $<$ 1 sec & 0 & 0 \\
            \quad \edsig{} & signature & 36 sec & $<$ 1 sec & 0 & 0 \\
            \quad \ecdsa{} (secp192r1) & signature & 880 sec & 7 sec & 105 & 10 \\
        \hline

        \multicolumn{4}{l}{\textbf{\mbedtls{}}~\cite{target-mbedtls} 3.3.0} \\
            \quad \aesecb{} & cipher & 2 sec & $<$ 1 sec & 68 & 68 \\
            \quad \aesgcm{} & aead-cipher & 4 sec & $<$ 1 sec & 216 & 76 \\
            \quad \chacha{} & aead-cipher & 7 sec & $<$ 1 sec & 0 & 0 \\
        \hline
        
        \multicolumn{4}{l}{\textbf{\openssl{}}~\cite{target-openssl} 3.0.0} \\
            \quad \aesecb{} & cipher & 115 sec & $<$ 1 sec & 52 & 52 \\
            \quad \aesgcm{} & aead-cipher & 117 sec & $<$ 1 sec & 166 & 60 \\
            \quad \chacha{} & aead-cipher & 117 sec & $<$ 1 sec & 0 & 0 \\
            \quad \edsig{} & signature & 556 sec & 4 sec & 0 & 0 \\
            \quad \ecdsa{} (secp192r1) & signature & 3128 sec & 30 sec & 1647 & 284 \\
        \hline

        \multicolumn{4}{l}{\textbf{\nettle{}}~\cite{target-nettle} 3.8.1 with \gmp{}~\cite{target-gmp} 6.2.1} \\
            \quad \aesecb{} & cipher & 2 sec & $<$ 1 sec & 32 & 32 \\
            \quad \aesgcm{} & aead-cipher & 3 sec & $<$ 1 sec & 108 & 40 \\
            \quad \chacha{} & aead-cipher & 2 sec & $<$ 1 sec & 0 & 0 \\
            \quad \edsig{} & signature & 104 sec & 4 sec & 0 & 0 \\
        \hline

        \multicolumn{4}{l}{\textbf{\scl{} - SiFive Cryptographic Library}~\cite{target-scl-metal} 20.08.00} \\
            \quad \ecdsa{} (secp256r1) & signature &  102 sec &  $<$ 1 sec & 5 & 2 \\
        \hline

        \multicolumn{4}{l}{\textbf{\libsodium{}}~\cite{target-libsodium} 1.0.18} \\
            \quad \chacha{} & aead-cipher & 2 sec & $<$ 1 sec & 0 & 0 \\
            \quad \edsig{} & signature & 12 sec & $<$ 1 sec & 0 & 0 \\
        \hline

    \end{tabular}
\end{table}